\documentclass[fleqn,twoside]{article}
\usepackage[headings]{espcrc2}

\usepackage{graphicx}
\usepackage{cite}

\newcommand{\tb}{\tan\beta}

\newcommand{\mA}{m_{A^0}}

\newcommand{\mg}{m_{\tilde{g}}}

\newcommand{\mb}{m_b}
\newcommand{\mt}{m_t}

\newcommand{\sur}{\ensuremath{\tilde{u}_{R}}}
\newcommand{\sdl}{\ensuremath{\tilde{d}_{L}}}
\newcommand{\sdr}{\ensuremath{\tilde{d}_{R}}}
\newcommand{\msdl}{\ensuremath{m_{\sdl}}}
\newcommand{\msdr}{\ensuremath{m_{\sdr}}}

\newcommand{\msur}{\ensuremath{m_{\sur}}}

\newcommand{\msusy}{\ensuremath{M_{\mathrm{SUSY}}}}
\newcommand{\deltatt}{\ensuremath{\delta_{23}}}
\newcommand{\hqq}{\ensuremath{h\to q\,{q'}}}
\newcommand{\hbs}{\ensuremath{h\to b\,s}}
\newcommand{\htc}{\ensuremath{h\to t\,c}}
\newcommand{\sigmapphqq}{\ensuremath{\sigma(pp\to \hqq)}}

\newcommand{\sigmapphbs}{\ensuremath{\sigma(pp\to \hbs)}}

\newcommand{\sigmapphtc}{\ensuremath{\sigma(pp\to \htc)}}

\newcommand{\Ghbs}{\ensuremath{\Gamma(h\to q\,{q'})}}
\newcommand{\bsg}{\ensuremath{b\to s\gamma}}
\newcommand{\Bbsg}{\ensuremath{B(b\to s\gamma)}}
\newcommand{\GeV}{\mbox{\,GeV}}
\newcommand{\squark}{\ensuremath{\tilde{q}}}

\newcommand{\pb}{\mbox{\,pb}}
\newcommand{\fb}{\mbox{\,fb}}

\newcommand\pubblock{\rightline{\texttt{\pubnumber, \hepnumber}}}
\newcommand\pubnumber{UAB-FT-597, UB-ECM-PF-06/03}

\newcommand\hepnumber{hep-ph/0601191}

\newcommand{\BejarJHEP}{Bejar:2004rz}

\newcommand{\PDG}{Eidelman:2004wy}

\begin{document}


\title{SUSY Higgs boson flavor-changing neutral currents at the LHC}
\author{Santi
  B{\'e}jar\address{Grup de F{\'\i}sica Te{\`o}rica,
    Universitat Aut{\`o}noma de Barcelona,
    E-08193, Bellaterra,
    Barcelona, Catalonia, Spain}\address[IFAE]{Institut de F{\'\i}sica d'Altes Energies, Universitat
    Aut{\`o}noma de Barcelona,
    E-08193, Bellaterra, Barcelona, Catalonia, Spain},
 \underline{Jaume Guasch}\address[ECM]{HEP Group, Departament d'Estructura i Constituents de la
    Mat{\`e}ria, Universitat de Barcelona,\\
    Diagonal 647, E-08028 Barcelona, Catalonia, Spain}, 
  Joan Sol{\`a}\addressmark[ECM]\addressmark[IFAE]%
\thanks{Contribution to the proceedings of \textit{7th
      International Symposium on Radiative Corrections},  Shonan
    Village, Japan, October 2-7, 2005}%
}
\runauthor{Santi
  B{\'e}jar,
 Jaume Guasch,
  Joan Sol{\`a}}

\begin{abstract}
\pubblock
We compute and analyze the Flavor-Changing Neutral Current (FCNC)
interactions of Minimal Supersymmetric Standard Model Higgs bosons
($h\equiv h^0,\ H^0,\ A^0$) with heavy quarks (top and bottom), focusing
on the strongly-interacting sector. We correlate the Higgs bosons
production cross-section at the LHC with the FCNC decay branching ratios
and find the maximum allowed values of the production rates,
$\sigma(pp\to h\to q{q}')\equiv\sigma(pp\to h) \times B(h\to q{q}')$
($qq'\equiv t c$ or $bs$) after taking into account limits from low energy
data on flavor-changing interactions. We single out the top quark channel,
with a maximum production rate of $\sigma^{\rm max}(pp\to h\to tc)\simeq
10^{-3}-10^{-2}\pb$, as the most promising FCNC channel to be detected at the
LHC.
\end{abstract}
\maketitle
\setcounter{footnote}{0}

\section{Introduction}
The possibility that Nature hides the last possible symmetry of
the $S$-matrix, Supersymmetry (SUSY), in the form of a broken
symmetry of the fundamental interactions is one of the most
intriguing puzzles over the last $30$ years of high energy
physics. SUSY offers a most tantalizing paradigm to unify
Particle Physics interactions with Gravity, and it might even
provide a clue to a more fundamental superstring unification of
all interactions. As such SUSY will be scrutinized in great
detail at the LHC. It also offers a possible solution to the
longstanding naturalness problem in the Higgs sector of the
SM~\cite{Haber:1985rc}. If SUSY is realized around the TeV scale,
the LHC experiments shall be able to directly produce the SUSY
particles for masses smaller than a few
TeV\,\cite{Weiglein:2004hnDegrassi:2004ed}. Among the many
strategies devised to hunt for SUSY, Flavor-Changing Neutral
Currents (FCNC) offer perhaps a unique laboratory to seek for the
new signs of physics beyond the SM. In the SM the FCNC effects are
completely absent at the tree-level. At one loop, however, they
are possible, but then the contributions from the new particles
enter on equal footing with those from SM particles. In certain
regions of the parameter space the new physical effects may well
dominate the SM contributions. This is particularly so when the
SM one-loop effects turn out to be highly suppressed. In such
situations the sole observation of these FCNC processes would be
instant evidence of new physics. An example of this kind of
appealing scenarios occurs within the FCNC physics of the SM
Higgs boson ($H_{\rm SM}$) interactions with quarks. For example,
the FCNC vertex $H_{\rm SM}tc$ may lead (at one loop) to such
rare decays as $t\rightarrow H_{\rm SM}\,c$ or $H_{\rm
SM}\rightarrow t\,c$, depending on the mass of $H_{\rm SM}$. Both
of these modes are extremely suppressed at one loop, with
branching ratios of order $10^{-14}$ or
less~\cite{HtcSM,\BejarJHEP}, hence
$10$ orders of magnitude below other more conventional (and
relatively well measured) FCNC processes like $b\rightarrow
s\gamma$\,\cite{\PDG}.

The Minimal Supersymmetric Standard Model (MSSM) introduces new sources
of FCNC interactions mediated by the strongly-interacting
sector\footnote{For a brief description of these interactions see
e.g. Ref.~\cite{Sola:2006npbpc}, for details see~\cite{Guasch:1999jp}.}. They are
produced by the misalignment of the quark mass matrix with the
squark mass matrix, and the main parameter characterizing these
interactions is the non-flavor-diagonal term in the squark-mass-matrix,
which we parameterize in the standard fashion\cite{GabbianiMisiak} as
$
  (M^2)_{ij} = \delta_{ij} \tilde{m}_i \tilde{m}_j\ (i \neq j)
$,
$\tilde{m}_i $ being the flavor-diagonal mass-term of the $i$-flavor
squark. Since there are squarks of different chiralities, there are
different $\delta_{ij}$ parameters for the different chirality
mixings. In this work we will assume flavor-mixing only among the
Left-Chiral squarks, since these mixing terms are expected to be the
largest ones by Renormalization Group analysis~\cite{Duncan:1983iq}. 

Some work in relation with the MSSM Higgs bosons FCNCs has already been
performed~\cite{Guasch:1999jp,\BejarJHEP,HfcncMSSM,Bejar:2005kv}.
In this work
we compute and analyze the production of any MSSM Higgs boson
($h=h^0,H^0,A^0$) at the LHC, followed by the one-loop FCNC decay $h\rightarrow
b\,s$ or $h\rightarrow t\,c$, and we find the maximum production rates
of the combined cross-section:
\begin{eqnarray}
  \sigmapphqq&
  \equiv&
  \sigma(pp\to h X)B(\hqq)\nonumber\\
      B(\hqq)&\equiv&\frac{\Gamma(h\to
          q\,\bar{q'}+\bar{q}\,q')}{\sum_i \Gamma(h\to X_i)}
    \label{eq:hqq-def}
\end{eqnarray}
$qq'$ being a pair of heavy quarks $(qq'\equiv bs$ or $tc)$, taking into
account the restrictions from the experimental determination of
$\Bbsg$~\cite{\PDG}. For other signals  of SUSY FCNC at the LHC,
without Higgs bosons couplings see Ref.~\cite{Sola:2006npbpc}.

\section{Computation setup}
In this section we give a summarized explanation of the computation. 
For further details see Refs.\cite{Bejar:2005kv,Bejar:2004rz}. 

We include the full one-loop SUSY-QCD contributions to the FCNC partial
  decay widths $\Ghbs$ in the observable (\ref{eq:hqq-def}).

 The Higgs sector parameters (masses and CP-even mixing angle
  $\alpha$) have been treated using the leading $\mt$ and $\mb\tb$
  approximation  to the one-loop result~\cite{Dabels}. 

 The Higgs bosons total decay widths $\Gamma(h\to X)$ are computed
  at leading order, including all the relevant channels: $\Gamma(h\to
  f\bar{f},ZZ,W^+W^-,gg)$. The off-shell decays
  $\Gamma(h\to ZZ^*,W^{\pm}W^{\mp*})$ have also been
  included. This is necessary to consistently compute the
  total decay width of $\Gamma(h^0\to X)$ in regions of the parameter
  space where the maximization of the cross-section (\ref{eq:hqq-def})
  is obtained at the expense of greatly diminishing the partial decay widths of the two-body
  process $h^0\rightarrow b\bar{b}$ (due to dramatic quantum effects that may reduce
  the CP-even mixing angle  $\alpha$ to small values~\cite{smallalphaeff}).
  The one-loop decay rate $\Gamma(h\to gg)$
  has been taken from~\cite{Spira:1995rr} and the off-shell decay
  partial widths have been recomputed explicitly.

 The MSSM Higgs boson production cross-sections  have been computed
using the programs  \texttt{HIGLU 2.101} and \texttt{PPHTT
1.1}\,\cite{HigluHqq,Spira:1995rr}. These
programs include the following channels: gluon-gluon fusion, 
and associated production with top-
and 
bottom-quarks. We have used the leading order
approximation for all channels. The QCD renormalization scale is
set to the default value for each program. We
have used the set of CTEQ4L PDF\cite{Lai:1996mg}.

For the constraints on the FCNC parameters, we use $\Bbsg=(2.1-4.5)\times
10^{-4}$ as the experimentally allowed range within three standard
deviations~\cite{\PDG}. We also require that the sign of the $\bsg$
amplitude is the same as in the SM~\cite{Gambino:2004mv}\footnote{This
    constraint automatically excludes the \textit{fine-tuned}
    regions of Ref.~\cite{Bejar:2004rz}.}.

Running quark masses ($m_q(Q)$) and strong coupling constants
($\alpha_s(Q)$) are used throughout, with the renormalization scale set
to the decaying Higgs boson mass in the decay processes.

Given this setup, we have performed a Monte-Carlo
maximization~\cite{Brein:2004kh} of the
cross-section~(\ref{eq:hqq-def}) over the MSSM parameter space, keeping
the parameter $\tb$ fixed, and under the simplification that the squark
and gluino
soft-SUSY-breaking parameter masses are at the same scale:
\begin{equation}
\msdl=\msdr=\msur=\mg\equiv\msusy\ \ \ .
\label{eq:susyscale}
\end{equation}
\begin{figure}[t]
\resizebox{6cm}{!}{\includegraphics{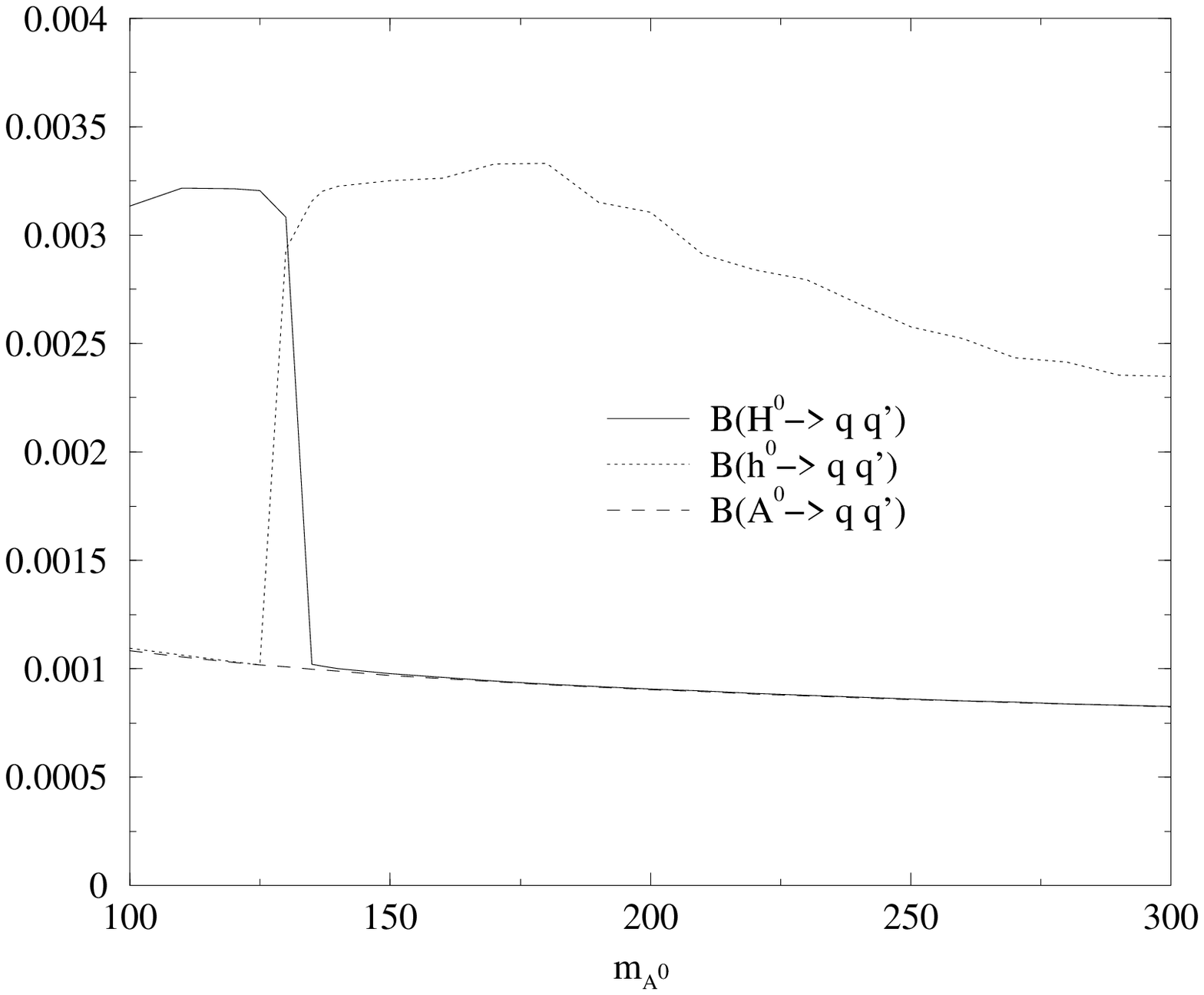}}
\caption{The maximum value of $B(h\to bs)$ as a function of $\mA$, for $\tb=50$.\label{fig:hbsma0}}
\end{figure}

Although we have performed a full one-loop computation, involving the
exact diagonalization of the $6\times6$ squark mass matrix, it is
enlightening to look at the approximate leading expressions to
understand the qualitative trend of the results. The SUSY-QCD contribution
to the $\bsg$ amplitude can be approximated to:
\begin{equation}\label{bsgamma}
A^{SQCD}(\bsg)\sim \deltatt\,\frac{m_b(\mu-A_b\tb)}{\msusy^2}\,,
\end{equation}
whereas the MSSM Higgs bosons FCNC effective couplings behave as: 
\begin{equation}
g_{h q\bar{q'}}   \sim \delta_{23} \frac{-\mu \,
    \mg}{M_{SUSY}^2} \left\{ \begin{array}{cc}
 \sin(\beta-\alpha_{\rm eff})& (H^0)\\
    \cos(\beta-\alpha_{\rm eff}) & (h^0)\\
 1& (A^0)
  \end{array}\right..
\label{eq:approxleading}
\end{equation}
The different structure of the amplitudes in eqs.~(\ref{bsgamma})
and~(\ref{eq:approxleading}), will permit to obtain and appreciable FCNC
Higgs boson decay rate, while the prediction for $\Bbsg$ stays inside
the experimentally allowed range.

\section{Bottom-strange channel}
\label{sec:hbs}

For the analysis of the bottom-strange production channel, we study
first the Higgs boson branching ratio (\ref{eq:hqq-def}).
Figure~\ref{fig:hbsma0} shows the maximum value of $B(h\to bs)$ 
as a function of the pseudoscalar Higgs boson mass $\mA$. We observe
that fairly large values of $B(h^0\to bs)\sim 0.3\%$ are
obtained. Table~\ref{tab:maximnowindow} shows the actual values of the
maximum branching ratios, and the parameters that provide them for each
Higgs boson. Let us discuss first the general trend, which is valid
for all studied processes: the maximum is attained at large $\msusy$ and
moderate $\deltatt$. The SUSY-QCD contribution to $\bsg$~(\ref{bsgamma})
decreases with $\msusy$, therefore to keep $\Bbsg$ in the allowed range
when $\msusy$ is small, it has to be compensated with a low value of
$\deltatt$, providing a small FCNC effective
coupling~(\ref{eq:approxleading}). On the other hand, at large $\msusy$
the second factor in eq.~(\ref{bsgamma}) decreases, allowing a larger
value of $\deltatt$. Thus, the first factor in
eq.~(\ref{eq:approxleading}) grows, but the second factor in
eq.~(\ref{eq:approxleading}) stays fixed (provided that $|\mu|\sim
\msusy$), overall providing a larger value of the effective coupling. 
On the other hand, a too large value of $\deltatt$ has to be compensated
by a small value of $|\mu|/\msusy$ in (\ref{bsgamma}), provoking a
reduction in (\ref{eq:approxleading}). In the end, the balance of
the various interactions involved produces the results of
Table~\ref{tab:maximnowindow}. 
\begin{table}[tb]
\rightline{\begin{tabular}{|c||c|c|c|} \hline $h$ &  $H^0$
& $h^0$ & $A^0$ \\
\hline\hline $B(h\to bs)$ &
$9.1\times 10^{-4}$ & $3.1\times 10^{-3}$ & $9.1\times 10^{-4}$\\
\hline $\Gamma(h\to
X)$ & $11.2 \GeV$ & $1.4\times 10^{-3} \GeV$ & $11.3 \GeV$
\\\hline $\delta_{23}$ & $10^{-0.43}$& $10^{-0.8}$ &
$10^{-0.43}$\\\hline \msusy & $1000 \GeV$ &  $975 \GeV$ &
$1000 \GeV$\\\hline $A_b$ & $-1500 \GeV$ & $-1500 \GeV$ & $-1500
\GeV$\\\hline $\mu$ & $-460 \GeV$ & $-1000 \GeV$ & $-460 \GeV$
\\\hline \Bbsg &  $4.49\times 10^{-4}$ &  $4.48\times 10^{-4}$
&$4.49\times 10^{-4}$ \\\hline
\end{tabular}}
\caption{Maximum values of $B(h\to bs)$ and corresponding SUSY
parameters for
  $\mA=200\GeV$,$\tb=50$.\label{tab:maximnowindow}}
\end{table}

The maximum value of the branching ratio for the lightest Higgs boson
channel is obtained in the \textit{small $\alpha_{\rm eff}$
  scenario}~\cite{smallalphaeff}. In this scenario the
coupling of bottom quarks to $h^0$ is extremely suppressed. The
large value of $B(h^0\to bs)$ is obtained because the total decay width
$\Gamma(h^0\to X)$ in the denominator of (\ref{eq:hqq-def}) tends to
zero (Table~\ref{tab:maximnowindow}), and not because a large FCNC
partial decay width in its 
numerator~\cite{Bejar:2004rz}. 

\begin{figure}[t]
        \includegraphics*[width=6cm]{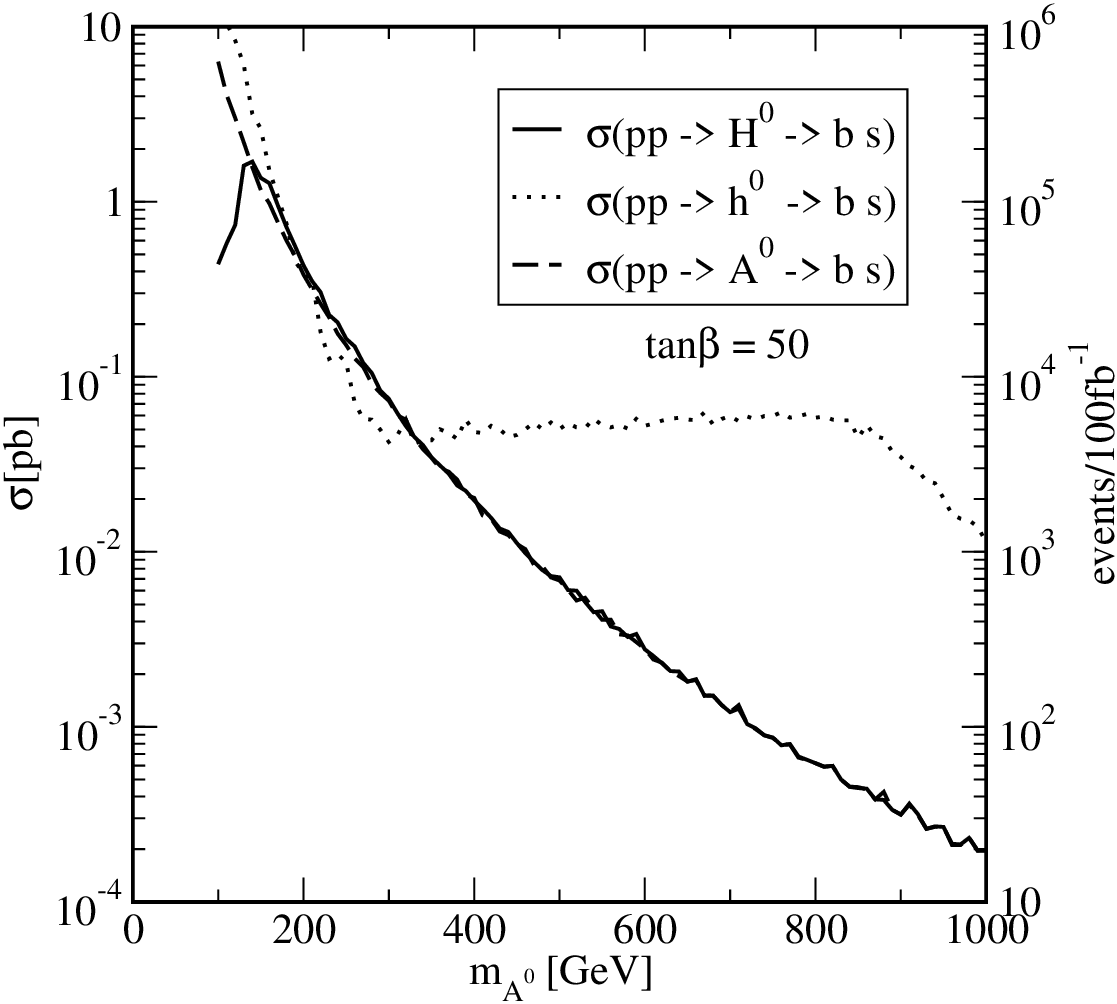} 
    \caption{Maximum SUSY-QCD contributions to $\sigma(pp\rightarrow
      h\rightarrow b\,s)$ as a function of
      $\mA$, $\tb=50$.}
    \label{fig:hbsprod}
\end{figure}
\begin{table}
    \rightline{
    \begin{tabular}{|c||c|c|c|}
        \hline
        $h$ &  $H^0$ & $h^0$ & $A^0$ \\\hline\hline
        \sigmapphbs &  $0.45\pb$ & $0.34\pb$ & $0.37\pb$ \\\hline
        events/$100\fb^{-1}$ & $4.5\times10^4$ & $3.4\times 10^4$ & $3.7\times10^4$\\\hline
        $B(h\to bs)$ & $9.3\times 10^{-4} $& $2.1\times 10^{-4} $& $8.9\times10^{-4} $ \\\hline
        $\Gamma(h\to X)$ & $10.9\GeV$ & $1.00\GeV$ & $11.3\GeV$
        \\\hline
        $\delta_{23}$ & $10^{-0.62}$ & $10^{-1.32}$ & $10^{-0.44}$ \\\hline
        $m_{\squark}$ & $990\GeV$ &  $670\GeV$ & $990\GeV$ \\\hline
        $A_b$ & $-2750\GeV$ & $-1960\GeV$ & $-2860\GeV$ \\\hline
        $\mu$ & $-720\GeV$ & $-990\GeV$ & $-460\GeV$ \\\hline
        \Bbsg & $4.50\times 10^{-4}$ & $4.47\times 10^{-4}$ & $4.39\times
        10^{-4}$ \\\hline
    \end{tabular}}
    \caption{Maximum value of $\sigmapphbs$ at the LHC,  for $\mA=200\GeV$ and
      $\tan\beta=50$. Shown are also the corresponding values of 
      $B(h\to bs)$ and of the total width of
      the Higgs bosons, together with the values of the SUSY
      parameters.}
    \label{tab:hbs-maxims}
\end{table}

The leading production channel of
$h^0$ at the LHC at high $\tb$ is the associated production with bottom quarks, and
therefore the $h^0$ production will be suppressed when $B(h^0\to bs)$ is
enhanced. We have to perform a combined analysis of the full
process~(\ref{eq:hqq-def}) to obtain the maximum production rate of FCNC
Higgs bosons meditated events at the LHC. Figure~\ref{fig:hbsprod} and
Table~\ref{tab:hbs-maxims} show the result of the maximization of the
production cross-section~(\ref{eq:hqq-def}). The central column of
Table~\ref{tab:hbs-maxims} shows that when performing the combined
maximization $\Gamma(h^0\to X)$ has a much larger value, and
therefore the maximum of the combined cross-section is not obtained in
the \textit{small $\alpha_{\rm eff}$ scenario}. The number of expected
events at the LHC is around 50,000 events/100$\fb^{-1}$. While it is a
large number, the huge $b$-quark background at the LHC will most likely
prevent its detection. Note, however, that the maximum FCNC branching
ratios are around $10^{-4}$--$10^{-3}$, which is 
 at the same level that the already measured $\Bbsg$.

\section{Top-charm channel}

\begin{figure}[tp]
        \includegraphics*[width=6cm]{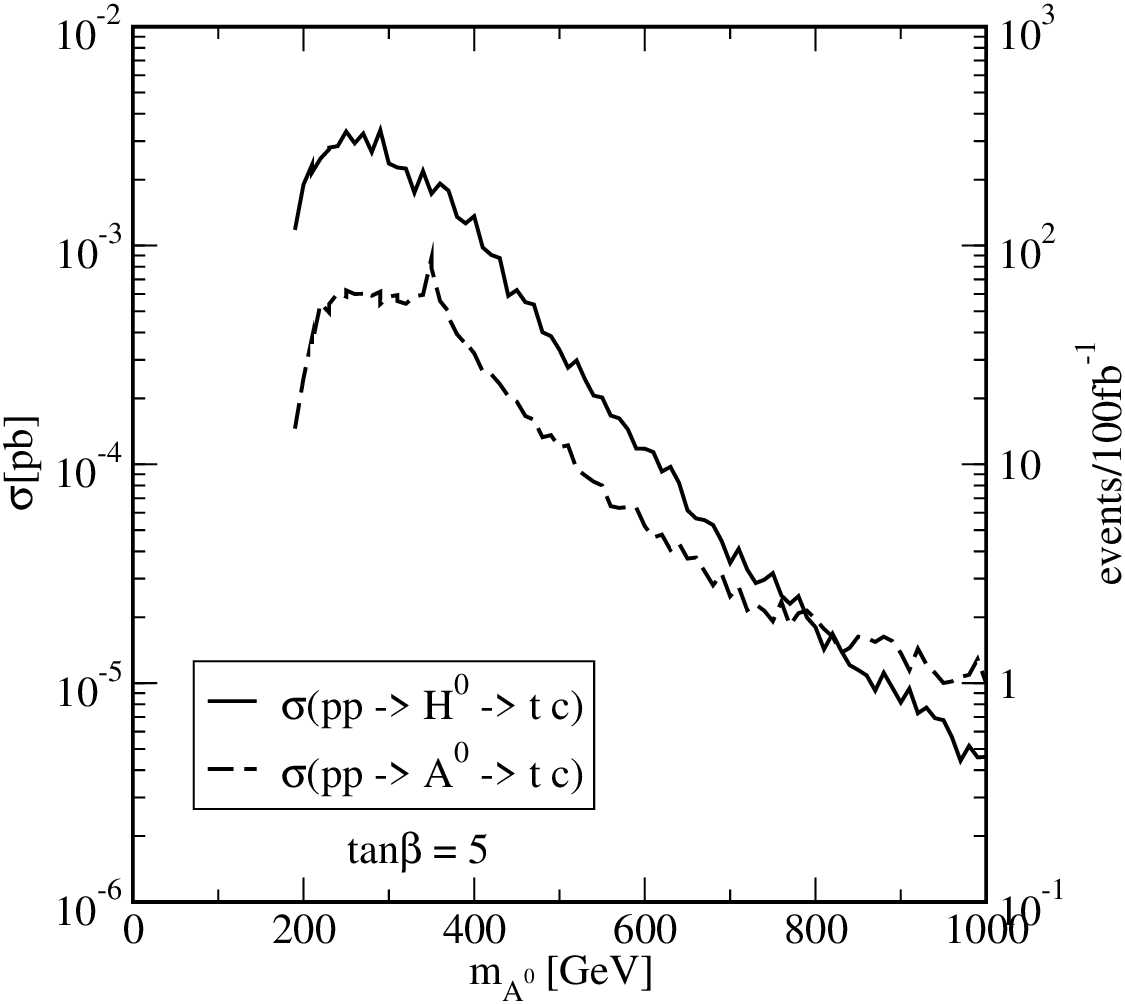}       
    \caption{Maximum SUSY-QCD contributions to $\sigma(pp\rightarrow
      h\rightarrow t\,c)$ as a function of $\mA$, $\tb$=5.}
    \label{fig:htcprod}
\end{figure}

The results of the numerical scan for this channel are similar to
the $bs$ channel, so we will focus mainly on the differences.
Figure~\ref{fig:htcprod} shows the maximum value of the
production cross-section $\sigmapphtc$ as a function of $\mA$, while
Table~\ref{tab:htc-maxims} shows the actual value of the maximum,
together with the SUSY parameters that provide them, for
$\mA=300\GeV$. Only the heavy neutral Higgs bosons contribute to this
channel. The general trend explained in section~\ref{sec:hbs} is also
valid here\footnote{Recall that the $\delta_{ij}$
  parameters in the up-sector are related to the corresponding
  parameters in the down-sector by the Cabibbo-Kobayashi-Maskawa matrix,
  see e.g.\cite{GabbianiMisiak}, and are therefore
  constrained by $\Bbsg$.}. 
From Table~\ref{tab:htc-maxims} we would expect some 300
events/100$\fb^{-1}$ at the LHC. However, the maximum is attained at low
$\tb$, and we have fixed a moderate value of $\tb=5$. By taking lower
values of $\tb$ we find that the number of events grows
fast\cite{Bejar:2005kv}, e.g. up to 
$\sim(500,900,2000)$ if we would have chosen $\tb=(4,3,2)$
respectively. Due to the single top quark signature the detection of this channel should
be feasible at the LHC. 

\section{Conclusions}

We have performed a computation of Higgs boson mediated FCNC events at
the LHC. Alternative effects of SUSY FCNC have been considered
in Ref.\cite{Sola:2006npbpc}. 
Note from eq.~(\ref{eq:approxleading}) that the present effects, though
smaller than the latter, do not
decouple with $\msusy=\mg=|\mu|$.
The maximum expected branching ratio for
the $bs$ 
channel is large, and we expect a maximum of $\sim 50,000$
events/100$\fb^{-1}$ in this channel at the LHC, which might be
difficult to detect due to the huge background. For the $tc$ channel we
expect a maximum of  $\sigma^{\rm max}(pp\to h\to tc)\simeq
10^{-3}-10^{-2}\pb$, which means several thousand events per
100$\fb^{-1}$ at the LHC. Due to the single top quark signature they
should be easier to detect than the $bs$ channel, providing the key to a
new door to study  physics beyond the Standard Model.
It is now an
experimental challenge to use this key to open the door, and prove that
these events can be effectively be separated from the background.

\textit{Acknowledgements}
The work of SB has been supported by CICYT (FPA2002-00648), by the EU
(HPRN-CT-2000-00152), and by DURSI (2001SGR-00188); JG  by a \textit{Ramon y
Cajal} contract from MEC (Spain); JG and JS in part by MEC and
FEDER (2004-04582-C02-01) and by DURSI (2005SGR00564).

\providecommand{\href}[2]{#2}

\end{document}